# USING MOBILE SERVICE FOR SUPPLY CHAIN MANAGEMENT: A SURVEY AND CHALLENGES

Selwa ELFIRDOUSSI[1]

[1] EMINES, School of Industrial Management, University Mohamed 6 Polytechnic, 43150, Benguerir, Morocco,

Selwa.elfirdoussi@emines.um6p.ma

## ABSTRACT

*Efficient supply chain management calls for robust analytical and optimal models to automate its process. Therefore, information technology is an essential ingredient that integrates these tools in supply chain. With the emergence of wireless, the high technologies and the reliability of mobile devices, mobile web services draw a promising horizon facing economic challenges. They offer new personalized services to each actor in the supply chain on their mobile devices at anytime and anywhere. This paper presents a literature review of mobile web service implemented on the industry context based on the supply chain management approach. First, a large definition of mobile web service and some proposal architecture are exposed. Then the paper discuss some generic related work on mobile web service focusing on supply chain management. Finally some challenges on m-service oriented supply chain management are proposed.*

## KEYWORDS

*Mobile web service; smart industry; Supply chain management; Service Oriented Architecture.*

## 1.INTRODUCTION

Efficient supply chain companies require a relevant management that is resuming on Supply Chain management (SCM). The SCM represents the concept and its effectiveness for practical application [1]. The objective of managing the supply chain is to synchronize the requirements of the customer with all the participants of the supply chain process, in order to effect a balance between what are often seen as conflicting goals of high customer service, low inventory management, and low unit cost [2].

The relevant key for SCM is the process communication, for that, it requires multiple implication of all the internal or external actors: employees, suppliers and customers to the company using business communications. Each person has to act and give the feedback in real-time and anywhere. In addition, the globalization has led companies to move their production factory, and to focus on the market of emerging countries.

Web service provides mechanisms to simplify data exchange by internet regardless of their implementation platforms. This technology has eliminated the complexity associated to data exchange from application, supporting different types of devices and different operating systems. It allows changes between data server with others, personal computer and mobile devices. Furthermore, due to the large amount of work in wireless, the high technologies and the mobile devices, using the mobile computer science have set peak some applications as mobile web service. In SCM, mobile web service technologies impose themselves by proposing the mobility of web service. It allows the deployment of new services to consumers in their mobile devices as phone, PDA or lap top, to be provided and consumed at anytime and anywhere [3]. As the traditional web service, mobile web service can also have a semantic description [4] to interact smartly between all participants.

This study, therefore, aims to provide a systematic and inclusive evaluation of research articles in order to gain insights into the applications of mobile service in supply chain. It also aims to develop a classification to analyse the extant literature in this subject area to provide a reference for researchers to maximize effort value in future research. Then, the paper list a proposed model for company's supply chain based on service task to coordinate theirs process.

The remainder of this paper is organized as follows: section II introduces a definition of mobile web service and presents some proposal architecture and their classification and exposes some relevant work in the mobile web service technology. Section III presents the concept of supply chain management and the research methodology proposed for ensuring it, then it focuses on the application proposed for mobile web service in the supply chain management face. The last section deals with the mobile web service oriented supply chain management by proposing an integrated model and some relevant research focus.

## 2. MOBILE WEB SERVICE

Web Service Technologies are the most adopted alternative for implementing the Service-Oriented Computing paradigm, in which developers can discover and combine specific functionality offered by third-parties instead of developing them. According to this technology, mobile web service is becoming the suitable technology to provide and consume service in the mobile environment. In fact, mobile has been the main device to work by integrating all the necessary functionalities using high technologies. In parallel, wireless communication achieve the main development to increase quality of transmission and improve its efficiency.

Therefore, the researchers were encouraged to integrate the web service technology in the mobile environment by proposing mobile web service with acceptable performance and with no impact on the regular use of mobile devices [5, 3]. In fact, Mobile web service is the technology that allows to provide, consume, web service using mobile devices such as phone, tablet, laptop or other wireless device, permitting users to access services at anytime and anywhere [6]. Furthermore, Mobile web service has become increasingly relevant because it has a number of benefits. It allows users to invoke and consume web service in their mobile phone and get desired information anytime and anywhere. It provides the response quickly according to the need of users and theirs queries. It increases the customer satisfaction and reduces the cost. It enables access to real-time information on mobile using the intranet or extranet enterprise.

Several criteria characterize any solution of the mobile web services such as scalability, availability, reliability, ubiquity, security, ease of use and immediacy [7]. Those criteria are fundamental to deploy mobile web service relayed to the mobile device and the wireless connection. Therefore, the researcher community was based to propose a mobile web service architecture service by taking into account the different characteristics [8].

### 2.1 Mobile web service : Architecture and classification

Current mobile Web service architectures are basically adapted to the traditional Web service approach by applying SOA (Service-oriented Architecture) Principles [9]. Therefore, a variety of mobile web service architectures were proposed by researchers regarding the Service-oriented Architecture (SOA) such as "proxy-based" or "Peer To Peer" or "Asymmetric" [3]. The proxy-based architecture is considered as the easiest approach to assure the efficiency of mobile web service deployment. In fact the proxy is usually a high-end machine attached to the fixed network [3].

To generalize the mobile web service architecture, the next figure resume a generic technical architecture assuring communication from mobile client and mobile service server presented in [10]. On the client-side, mobile devices can consume web service by the following ways:

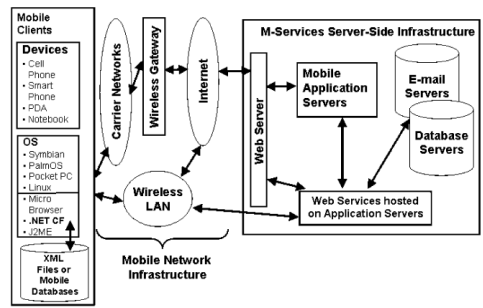

Figure 1. A generic Mobile Web Service Architecture [5].

Based on the generic mobile web service architecture, there are two ways to consume web service in this architecture. The first one is Mobile web clients that are hosted on the server application, and can be consumed on the server directly. The second one is mobile local client where the application is running on mobile devices by connecting to server using wireless or local area. In both cases, mobile devices invoke web service and return response to server using the appropriate protocol and format (XML, SOAP). On the server side, web services are deployed on application server and provide mobile devices with the functions to access data. In fact, the mobile and server can interact using internet. Some relevant works have proposed a new architecture for mobile web service [11, 12, 13, 14] that we expose in details later.

According to the generic architecture, mobile web service technology support in either connected or disconnected consummation mode by consumers or business applications [13]. Therefore, the mobile web service can be classified in two categories: target users (consumers vs. business) and network connection (disconnected vs. connected). The figure presents a matrix defining those categories. Data exchanged in the offline mode will be sent and synchronized with the server when mobile devices will be online and connected to the network area.

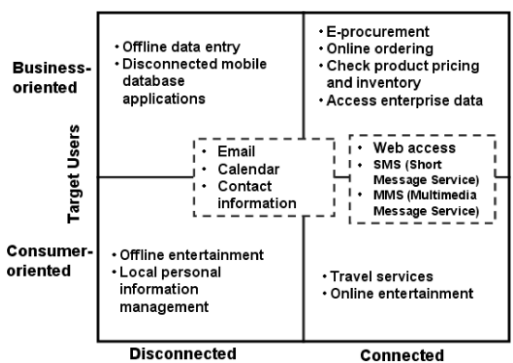

Figure 2. Mobile web service classification [13].

The main objective of the first category is to use the existent web service provided by internet in mobile device. The second category presents the providers and consumers with two mobile devises that communicate peer to peer [7].

## 2.2 Litterature review

With the fast integration of Mobile web service, the researchers were focalized on this technology proposing a 1) new architecture; 2) discovering approach; 3) adding semantic description, 4) Invoke approach; 5) providing proper quality of service (QoS), or 6) proposing model for security. In this section, we discuss some recent relevant papers proposed in mobile web service technology classified by the category of research. The main categories concern the proposition of new architecture or a new discovery approach taking into account the quality of service. The table 1 summarizes the list analysed by reference and year for each analytical axe.

Table 1 : List of Analysed paper by axes

| | | | Analytical Axes | | | | | | |
|---|---|---|---|---|---|---|---|---|---|
| Reference | year | Approach proposed | Architecture | Discovering | Semantic | Invoke | Composition | Qos | Security |
| 11 | 2005 | HandHeld Flexible Representation | | | | | | | |
| 12 | 2010 | Communication architecture | x | | | | | | |
| 13 | 2005 | M-service portail | x | | | | | | |
| 14 | 2009 | Extensible Messaging and Presence Protocol | x | x | | | | | |
| 15 | 2012 | Lean and flexible system architecture | x | x | | | | | |
| 16 | 2012 | MIM Server | x | x | | x | | | |
| 4 | 2014 | Semantic matchmaking model | x | | x | | | x | |
| 18 | 2007 | M-service mediation framework | | x | | | | | |
| 19 | 2014 | A novel cloud-based framework | | x | | | | x | |
| 20 | 2006 | New agent based framework | | x | | x | x | x | |
| 21 | 2008 | MWSMF | | | | | | x | |
| 22 | 2008 | Discovery in peer to peer networks | | x | | | | | |
| 23 | 2013 | Personalized M-service Discovery | | x | | | | x | |
| 24 | 2013 | Dynamic Discovery and Invocation of M-service | | x | | x | | | |
| 25 | 2014 | Extended Mobile Host Complex Web service | x | | | | | | |
| 26 | 2013 | Towards SOA for M-service Composition | | | | | x | | |
| 27 | 2004 | Security in mobile ad hoc networks | | | | | | | x |
| 28 | 2007 | Evaluation of M-service security | | | | | | | x |

The key element of a technology is its architecture, and in mobile web service, some research was focus on proposing a new architecture. The authors in [11] propose a new architecture supporting mobile web service called the Handheld Flexible Representation (HHFR) that is designed for an optimized and expandable communication in mobile web services. The architecture provides alternative representations other than XML-based SOAP and fast communication transport options. HHFR separates message contents from a SOAP message in Angle-Bracket syntax format, and service end-points in HHFR exchange separated-message contents in an optimized fashion. In the same context, Cobarzan presents [12] a communication architecture based XML exchange data from the mobile clients to an external middleware component. The middleware will act like a gateway that lightly communicates with the device in a client-server manner over a fast binary protocol and at the same time takes the responsibility of solving the request to the Web service. Chen et al. propose in [13] an m-service portal architecture which integrates mobile web services to provide adaptive and personalized services accommodating the constraints of mobile devices. The authors propose in [14] a new architecture for managing web service directories using XMPP. In the proposed architecture each mobile device may act as a service provider as well as a service consumer. zaplata presents in [15] a lean and flexible system architecture which supports both mobile web service consumers and providers. The architecture integrate multiple protocols that depends on their capabilities, it also allows a dynamically access suitable service at runtime. Artail et al. propose in [16] a proxy-based architecture for dynamic discovery and invocation of Web Services from mobile devices.

Web service discovery is a main challenge despite the enhanced proposed methods based on information retrieval techniques [17], and the problem concerns also Mobile web service. Some authors who propose an architecture of mobile web service have also integrated the discovery approach based on their architecture [15, 16, 14]. Elabd presents in [4] a semantic matchmaking model to discover the Web services taking into account the user requirements, Quality of Services (QoS) and Mobile Specification as required parameters. Their work concerns the degree of match calculation taking into account the QoS and Mobile specification. The authors in [18] propose some QoS and discovery research on mobile web service and a mediation framework for mobile Web service provisioning.

Algazzar introduces in [19] a cloud-based framework for mobile Web service discovery (DaaS) for resource constrained and mobile environments. DaaS has two main objectives: (1) pushing the resource-intensive processes of service discovery to the cloud in order to save the already scarce mobile resources, (2) incorporating the user preferences and context in mobile Web service discovery. The authors propose in [20] an agent-based mobile services framework using the wireless portal network, which eliminates XML processing on mobile clients. The framework contains an interface to select a service and a module to apply development and deployment for Web service providers. The authors address in [21] a proposition to reduce the delays by reducing the size of messages exchanged in mobile web service and [22] a concept of mobile web service provisioning and discovery in P2P networks. Elgazzar proposes in [23] a robust service discovery approach that makes use of user preferences and context information that mobile devices can use to personalize the discovery of mobile Web services. The authors in [24] propose an efficient and novel architecture for dynamic invocation and discovery of web services in mobile.

The approach introduced in [25] focuses on the architecture of the mobile cloud, which consists of a set of collaborative mobile devices that relies on an Extended Mobile Host Complex Web service Framework (EMHCWF). The authors have defined a mobile cloud computing framework that addresses a complex mobile web service mechanisms and allows distributed execution in a homogeneous manner.

A few works propose an approach for mobile service composition, in [26] the authors expose a solution to enable mobile Web service composition in an open, infrastructure-less environment based on loosely coupled SOA techniques.

Mobile devices are opened to all by communication cards such as Bluetooth, wireless, etc. Therefore, security is still the most important criteria in mobile environment, and has become a primary challenge in order to protect communication between mobiles [27]. The paper [28] summarizes a study on analyzing the security for mobile web services provisioning. The authors have included on their analysis the wired web security specifications to the cellular world. They conclude that web service provisioning in mobile networks is still a great challenge because the traditional networks are not always appropriate for the mobile environment. Furthermore, they recommend adding encryption solution at hardware.

## 3. MOBILE WEB SERVICE FOR SUPPLY CHAIN MANAGEMENT

### 3.1 Supply chain management

Supply Chain Management (SCM) consists on the integration of companies' processes according the exchanges between the suppliers, the customers and the company. It includes, fundamentally, the purchasing, production and distribution processes. The figure 3 presents a summarized illustration of a company's supply chain based on user task. The concept of supply chain Management (SCM) has raised the prominences over the past ten years for many reasons [1]. The globalization of supply and the requirements of customers have encouraged companies to implement efficient solutions for the best coordination and automation of their process. This global orientation and the increased performance-based competition combined with the technology evolution and economic conditions have contributed sequentially on the marketplace. For that, availability and flexibility of the supply chain is required to keep their place in the market.

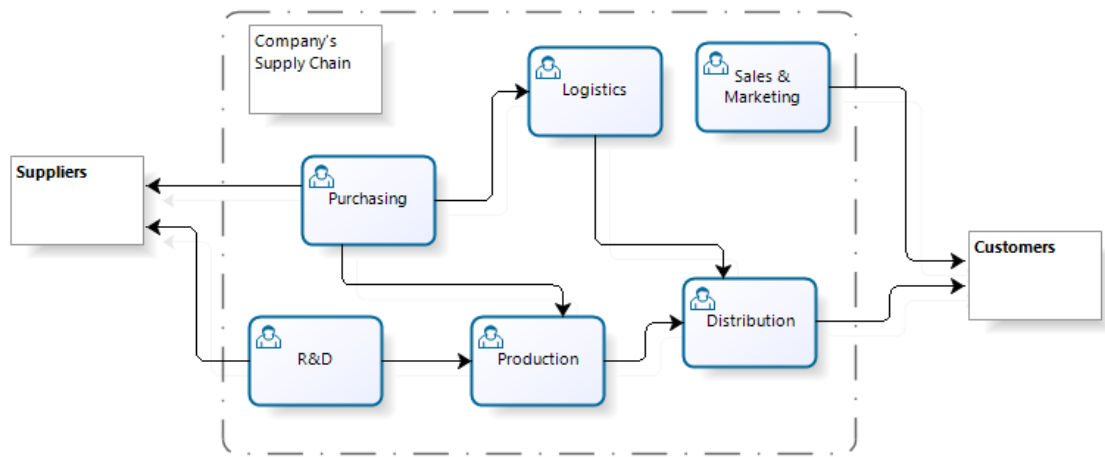

Figure 3. An illustration of a company's supply chain management based on user task

Since the concept of supply chain management was introduced, it has become a popular operations paradigm. This has intensified with the development of information and communication technologies (ICT) that include electronic data interchange (EDI), the Internet and World Wide Web (WWW) [29]. The complexity of SCM has also forced companies to go for online communication systems. Especially, with the implementation of applications based on mobile devices such as mobile web service.

### 3.2 Research methodology

The research methodology employed for developing the mobile web service for the successful application of mobile web service in SCM is the literature survey. We have collected literature primary through journals that are in the areas of operations management, supply chain, operations research and information systems. It was aimed at primarily helping researchers and practitioners in implementing a successful mobile web service for achieving an effective SCM. With this in mind, we looked at the literature that deals with mobile application SCM.

For the literature review related to m-service in SCM, two major categories have emerged: a) the supply management process [30]; b) the m-service analytical axes. For each category, we have listed some criteria to analyze the paper. In the first category, we have listed a majority of the SCM presented in sub section 3. In fact the m-service proposed has to cover the optimization of the process to ensure an effective SCM. In the second category, we have improved the criteria proposed in the literature review of section 2.

### 3.3 Litterature review

The advancement of mobile devices and communication technologies has been of help and time saving to the company. One major advantage of mobile phone communication in industry business is that data can be transmitted on real-time virtually worldwide [31]. Despite this, the development of mobile services for intra-organizational use is still rare and slow. The OCAD University has published a report [32] concerning the Mobile Innovation in service and its industrial applications. The team explains the important role that mobile technologies play in increasing the competitively of industrial companies. For this purpose, companies must focus on acquiring and developingmarketing skills using mobile web service. They should explore avenues to obtain additional real-time data and information to inform their decision-maker.

In this subsection, we discuss some related works for the proposed mobile web service integrated SCM based on the research methodology presented before. We focus the analysis on the process covered, mainly in the additional set of papers of table 2.

Table 2: The analyzed paper on m-service in SCM

| Reference | year | SCM Process | | | | | | | M-service analytic axes | | | | | | |
|---|---|---|---|---|---|---|---|---|---|---|---|---|---|---|---|
| | | Purchasing | R&D | Production | Sales & Marketing | Logistics | Inventory | Finance | Architecture | Discovering | Semantic | Invoke | Composition | Qos | Security |
| 33 | 2006 | x | | x | | x | | | | | | | | | |
| 34 | 2003 | | | x | | x | | | | x | | x | | | x |
| 35 | 2010 | | | | | x | | | | x | | | | | |
| 36 | 2012 | | | | | x | | | | x | | | | | |
| 37 | 2005 | | | | | x | | | | | | | | | |
| 38 | 2007 | | | x | | x | x | | | x | | | | | |
| 39 | 2016 | x | | x | | x | | | | | | | | x | |

Irena & et. propose a study in [33] about potentials of wireless technology for logistics and expose five criteria of suitability of logistics processes for mobile business. It concerns: Necessity to transport good, Necessity to perform task on site, Necessity of the real-time access to new data, Necessity of wireless data entry, Possibility for paperless workflow. They introduce many possibilities for mobile services, which will improve and optimize the underlying processes.

The authors has developed in [34] a mobile medical application that can be executed on a mobile platform. The application area allows a patient to be remotely monitored, while allowing the patient to increase its mobility. Currently, the software developed aims for real time monitoring of the heart rate and blood oxygen saturation. For enhanced functionality, additional features could easily be deployed e.g. data storage or data processing locally on the EIS sensor platform.

The qualitative research case study proposed in [35] has enabled the research team to gain an understanding of the challenges involved in a global logistics operation and to interpret them in a real-life setting. Furthermore, through this empirical study the benefits of using GPS vehicle tracking and integration portal technologies within the context of the transport container industry have been demonstrated.

The approach described in [36] concerns the e-commerce applications from web services. The applications implemented in heterogeneous platforms and areas can collaborate through Web Services. Presentation layer communicates with database systems through services interface layer. The objective is to get date and information for each company, which have involved the integrated application. The functionality application for e-logistics is available as a mobile web service and the platform is integrated with the GPS/GIS technology to assure an efficient logistics distribution management for tracking purpose.

In [37], Angeles focuses on the application of the technology of radio frequency identification (RFID) in logistical operations. Using the RFID technologies, Kelepouris et al. propose in [38] a tractability system on the food-industry.

The authors propose in [39] an integrated framework that includes five aspects of SCM, i.e. material flow and supply management, real-time information sharing and communication, coordination and integration in CSC, technology support for M-Internet, and associated safety issues.

### 3.4 Challenges and opportunities

When preparing this study, we concluded that the works proposed in mobile web service in SCM seems to be still in its infancy. Table 2, lists and compares multiple works, in the literature review, focused on mobile web service for supply chain management criteria. This analysis allows us to identify the main challenges required for mobile web service to implement an effective SCM. Based on Stuart Taylor from cisco has proposed in [40] some perspectives on the future of the mobile industry. And in order to enhance competitive advantage through effective SCM, and with the growing of the web technology and mobile communication, we propose a model for a multi agent based services, illustrated on figure 4, according to the company's schema management (Figure 3). The model consists on deploying for each process an m-service able to be interacted with the other. Under this model, we propose to integrate in the internal infrastructure a:

- Monitoring of the supply chain devices using m-service.
- Proposing an interactive composition of m-service according to the supply chain workflow.

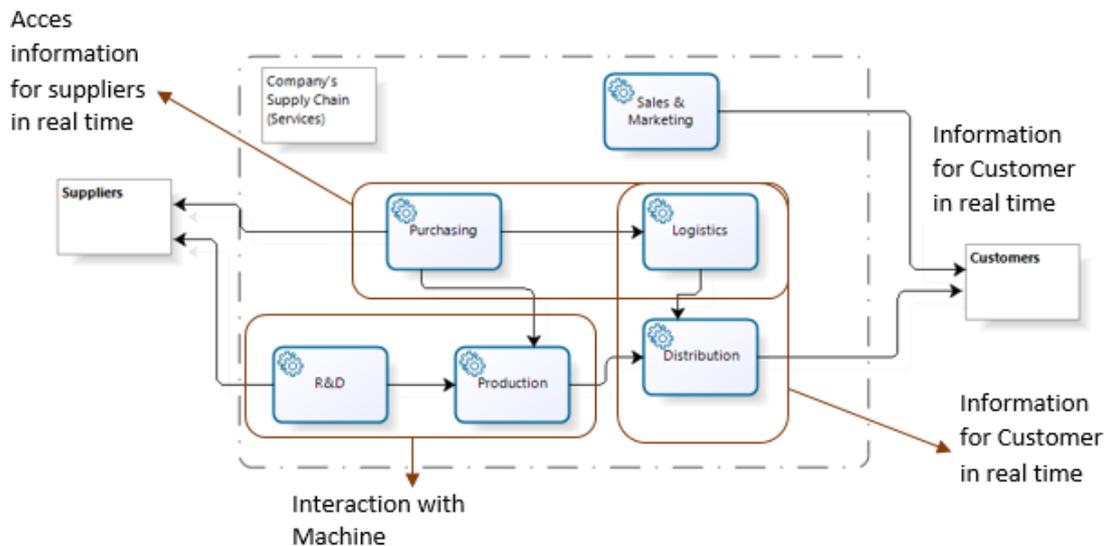

Figure 4. A proposed model for company's supply chain based on service task

In addition of the model, the m-services deployed have to ensure the analytic axes as defined in section 2. Therefore, we present below a list of challenges concerning the m-service, in order to automate the supply chain management completely:

- Mobile web service implemented on supply chain will require capabilities of service discovery, control, composition and quality of service to be registered in a public registry.
- Mobile web service implementation has to allow information access to all the supply chain processes in real-time, and sharing it with all the participants.
- Mobile web service for the User Interaction and the Machine Interaction such as production line, scanner devices, etc.
- Mobile web service implemented for an automation of the supply chain process with a controlling system.
- Mobile web service implemented has to ensure the quality of service according to the supply chain management.

Consequently, there are many challenges that can be exploited to propose an m-service oriented supply chain management mainly in two contexts. The first context consists on the integration of the classic web service proposed in SCM and the implementation of a mobile version. The second context is the development of a new m-service according to its generic architecture [5] for the automation of SCM process. A composition can also be proposed to get the process interaction for ensuring the SCM.

## 4. CONCLUSION AND FUTURE WORK

Researchers in Information Technology (IT) focalize their efforts to promote many SCM business solutions, ranging from enterprise resource planning (ERP), best of breed supply chain software, e-business applications to web services. So, according the SCM interacting environment of business, the systems proposed need to be interoperable. Therefore, mobile web service-based business information systems have developed a new paradigm of technology revolution in heterogeneous data source connectivity along supply chain. However, the increasing use of web services has raised new research challenges. To prove that, we have addressed, in this paper, a review of mobile web service in supply chain management. First, we propose a large definition of mobile web service and their architecture and classification. Then, we present some research of mobile web service for supply chain management. Unfortunately, we conclude that the development of mobile web services for intra-organizational use is still slow. The study proposed in this paper, encourages us to extend the research exposed in [41] to Mobile web service focusing on supply chain management for automating its process. The idea consists on tracking the proposed web service on supply chain management to be exploited in DIVISE framework and implement the mobile version. The composition module proposed in [42] will be used to compose some existing services according to their popularity and generate web service that can be consumed in mobile.

## REFERENCES

[1] J. T. Mentzer, W. DeWitt, J. S. Keebler, S. Min, N. W. Nix, C. D. Smith et Z. G. Zacharia, «Defining Supply Chain Mangement,» *Journal of Business logistics,* vol. 22, n° %12, pp. 1-25, 2001.

[2] S. c. Graham, «Integrating the supply chain,» *International Journal of Physical Distribution & Materials Management,* vol. 19, n° %18, pp. 3-8, 1989.

[3] K. Elgazzar, P. Martin et H. Hassanein, «Mobile Web Services: State of the Art and Challenges,» *International Journal of Advanced Computer Science and Applications,* vol. 5, n° %13, pp. 173-188, 2014.

[4] E. Elabd, P. M. El-Kafrawy et E. M. Abou-Nassar, «Semantic-based Web Services Selection for Mobile,» chez *The 9th International Conference on INFOrmatics and Systems*, 2014.

[5] S. Berger, S. Mcfaddin, C. Narayanaswami et M. Raghunath, «Web services on mobile devices-implementation and experience,» *Mobile Computing Systems and Applications,* pp. 100-109, 2003.

[6] F. AlShahwan et K. Moessner, «Providing SOAP Web services and RESTful Web services from mobile hosts,» chez *Internet and Web Applications and Services*, 2010.

[7] S. Hamida, Une approche basée agent mobile pour le m-service web sémantique, BISKRA: Université Mohamed Khider, 2014.

[8] T. M. Voigt, T. Naumowicz, H. Ritter et J. Schiller, «Performance considerations for mobile web services,» *Computer Communications,* p. 1097–1105, 2007.

[9] M. Treiber, C. Scherling et S. Dustdar, «Applying SOA Principles on Mobile Platforms,» chez *International Conference on Software Engineering*, Honolulu, 2010.

[10] F. Hirsch, J. Kemp et J. Ilkka, Mobile web services: architecture and implementation, John Wiley & Sons, 2007.

[11] S. Oh et G. C. Fox, «HHFR: A new architecture for Mobile Web Services Principles and Implementations,» Technical paper, 2005.

[12] A. COBÂRZAN, «Consuming Web Services on Mobile Platforms,» *Informatica Economică,* vol. 14, n° %13, pp. 98-105, 2010.

[13] M. Chen, D. Zhang et L. Zhou, «Providing web services to mobile users: the architecture design of an m-service portal,» *Int. J. Mobile Communications,* vol. 3, n° %11, pp. 1-18, 2005.

[14] S. Abhishek et V. Rohit, «A novel web service directory framework for mobile environments,» chez *International conferance on Web Services (ICWS)*, Anchorage, 2014.

[15] V. D. a. W. L. Sonja Zaplata, «Realizing Mobile Web Services for Dynamic Applications,» *AIS Transactions on Enterprise Systems,* vol. 2, n° %11, pp. 3-12, 2009.

[16] H. Artail, K. Fawaz et A. Ghandour, «A Proxy-Based architecture for dynamic discovery and invocation of Web Services from mobile devices,» *IEEE Transactions on Services Computing,* vol. 5, n° %11, pp. 99-115, 2012.

[17] S. Elfrirdoussi, Z. Jarir et M. Quafafou, «Popularity based Web Service Search,» chez *19th International Conference on web service*, Hawai, 2012.

[18] S. N. Srirama, M. Jarke et W. Prinz, «Mobile web services mediation framework,» chez *ACM/IFIP/USENIX International Middleware Conference*, New York, 2007.

[19] K. Elgazzar, H. Hassanein et P. Martin, «Cloud-based mobile Web service discovery,» *Pervasive and Mobile Computing,* vol. 13, p. 67–84, 2014.

[20] M. Adacal et A. Bener, «Mobile Web services: a new agent-based framework,» *IEEE Internet Computing,* vol. 10, n° %13, pp. 58-65, 2006.

[21] S. N. Srirama, M. Jarke et W. Prinz, «MWSMF: A Mediation Framework Realizing Scalable Mobile Web Service Provisioning,» chez *International conference on MOBILe Wireless MiddleWARE, Operating Systems, and Applications*, 2008.

[22] S. N. Srirama, M. Jarke, H. Zhu et H. Prinz, «Scalable mobile web service discovery in peer to peer networks,» chez *Internet and Web Applications and Services*, 2008.

[23] K. Elgazzar, P. Martin et H. Hassanein, «Personalized Mobile Web Service Discovery,» chez *World Congress on Services*, Santa Clara Marriott, CA, USA, 2013.

[24] R. Kumar et V. K. Kiran, «Dynamic Discovery and Invocation of Web services Through Android Mobile,» *International Journal of Engineering Trends and Technology,* vol. 4, n° %17, pp. 3182-3185, 2013.

[25] F. AlShahwan et M. Faisal, «Mobile cloud computing for providing complex mobile web services,» chez *Mobile Cloud Computing, Services, and Engineering*, San Francisco, 2014.

[26] C. Chang et S. Ling, «Towards an Infrastructure-less SOA for Mobile Web Service Composition,» chez *arXiv preprint arXiv:*, 1304.5045.

[27] H. YANG, H. LUO et F. YE, «Security in mobile ad hoc networks: challenges and solutions,» *IEEE wireless communications,* pp. 38-47, 2004.

[28] S. N. Rirama, M. Jarke et Prinz, «A performance evaluation of mobile web services security,» chez *3rd International Conference on Web Information Systems and Technologies*, 2007.

[29] A. Gunasekaran et E. W. Ngai, «Information systems in supply chain integration and management,» *European Journal of Operational Research,* vol. 159, n° %12, pp. 269-295, 2004.

[30] K. Croxton , S. Garcia-Dastugue et D. M. Lambert, «The Supply Chain Management Processes,» *The International Journal of Logistics Management,* vol. 12, n° %12, pp. 13-36, 2001.

[31] T. L. Rakestraw, R. V. Eunni et R. R. Kasuganti, «The mobile apps industry: A case study,» *Journal of Business Cases and Applications,* vol. 9, n° %11, pp. 1-26, 2013.

[32] A. Light, S. Punnett et K. Webb, Mobile Innovation: Ontario's Growing Mobile Content, Services, and Applications, Toronto: Mobile Experience Innovation Centre, 2012.

[33] O. Irena et R. Knut, «Mobile Business in Logistics,» *In GI Jahrestagung,* pp. 530-536, 2006.

[34] Ö. Åke, S. Linus, L. Per et D. Jerker, «Mobile Medical Applications Made Feasible Through Use of EIS Platforms,» chez *Instrumentation and Measurement Technology Conference*, USA, 2003.

[35] R. Michaelides, Z. Michaelides et D. Nicolaou, «Optimisation of logistics operations using GPS technology solutions: a case study,» chez *POMS conference*, Vancouver, 2010.

[36] K. S. Chinchu et G. Selvakumar, «Web Service based e-logistics application,» *International Journal of Modern Engineering Research,* pp. 084-088, 2012.

[37] R. Angeles, «supply-chain applications and implementation issues,» *Information systems management,* vol. 22, n° %11, pp. 51-65, 2005.

[38] T. Kelepouris, K. Pramatari et G. Doukidis, «RFID-enabled traceability in the food supply chain,» *Industrial Management & Data Systems,* vol. 107, n° %12, pp. 183-200, 2007.

[39] Q. Shi et X. Ding, «Mobile Internet based construction supply chain management: A critical review,» *Automation in Construction.,* 2016.

[40] S. Taylor, «The New Mobile World Order : Perspectives on the Future of the Mobile Industry,» Cisco, 2012.

[41] S. Elfrirdoussi, Z. Jarir et M. Quafafou, «Discovery and Visual Interactive WS Engine based on popularity: Architecture and Implementation,» *International Journal of Software Engineering and Its Applications,* pp. 213-228, 2014.

[42] S. Elfrirdoussi, Z. Jarir et M. Quafafou, «Web Service Composition Based on Popularity,» chez *CS & IT CSCP*, 2014.

**Authors**

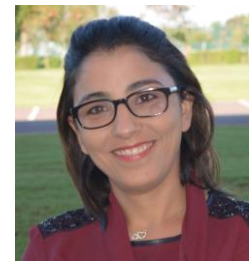

After obtaining a Degree as a computer science engineer from ENSIAS in 2000, I started my career as a consultant in the field of information system. In 2015, I obtained a PhD in Computer Engineering and Applied Mathematics from the Cadi Ayyad University of Marrakech in co-supervision with the Saint-Jérôme University of Marseille on the classification, discovery and composition of web services Based on popularity. Then, I joined the UM6P as a university professor at EMINES School of Industrial Management, where I am in charge of the computer science courses of the Integrated Preparatory Cycle and the Engineering Cycle. I am also a member of the EMINES industrial management research team. My research works focuses on the development of decision support systems for the advanced management of supply chains Integrating the Service Oriented Architecture (SOA).